\documentclass[pra,aps,preprintnumbers,superscriptaddress]{revtex4}
\usepackage{epsfig,amssymb,amsfonts}
\newcommand{\ket}[1]{\left | \, #1 \right\rangle}

\begin{document}
\title{Many body effects and cluster state quantum computation in strongly interacting
systems of photons}

\author{Dimitris G. Angelakis}
\affiliation{Science Department, Technical University of Crete,
Chania, Crete, Greece, 73100}
\author{Sougato Bose}
\affiliation{Department of Physics and Astronomy, University College
London, Gower St., London WC1E 6BT, UK}
\author{Alastair Kay}
\affiliation{Centre for Quantum Computation, Department of Applied
Mathematics
 and Theoretical Physics, University of Cambridge, Wilberforce Road, CB3 0WA, UK}
\author{Marcelo F. Santos}
\affiliation{Dept. de F\'{\i}sica, Universidade Federal de Minas
Gerais, Belo Horizonte, 30161-970, MG, Brazil}


\begin{abstract} We discuss the basic properties of a recently
proposed hybrid light-matter system of strongly interacting
photons in an array of coupled cavities each doped with a single
two level system. Using the non-linearity generated from the
photon blockade effect, we predict strong correlations
between the hopping photons in the array, and show the possibility of
observing a phase transition from a polaritonic insulator to a
superfluid of photons. In the Mott phase, this interaction can
be mapped to an array of spins. We show how the remaining Hamiltonian,
in conjunction with individual spin manipulation, can thus be used for
simulating spin chains (useful for state transfer protocols) and cluster
state quantum computation.
\end{abstract}

\maketitle
\bibliographystyle{apsrev}

\section {Introduction}

The intractability of simulating coherent many-body phenomena on a
classical computer is a major barrier to our understanding of many
condensed-matter systems and their dynamics, such as Mott to superfluid
phase transitions, high-temperature superconductivity, and
anti-ferromagnetism. One of Feynman's great insights was that this
problem could be overcome by using other quantum
systems, over which we have a much greater degree of control, as simulators.

Cold atoms in optical lattices have been one of the most successful
quantum simulators so far
\cite{fisher,jaksch,greiner,kasevich,Duan-Lukin-Demler-PRL}. However
it is still extremely interesting to explore which other systems
permit such phases and simulations, especially as the problem of
accessibility of the individual sites has been extremely difficult
to address.
Here, we review the properties of a recently proposed system which
consists of an array of coupled cavities, each doped with a single
two-level system. Coupled cavities arrays (CCAs) have been initially
proposed for the implementation of quantum gates \cite{ASYE_PLA_07}.
In \cite{ASB_PRA_07} we showed  that the atom-cavity interaction
could induce a non-linear interaction, commonly described as the
photon blockade effect \cite{blockade,blockade1}, enabling the
prediction of strong correlations between the hopping photons in the
array. Similarly to optical lattices, which demonstrate a superfluid
to Mott insulator phase transition \cite{jaksch},  a phase
transition was predicted between a superfluid of photons and a Mott
phase of hybrid light-matter excitations known as polaritons.
Simultaneously and independently with  \cite{ASB_PRA_07}, a similar
study for strongly interacting polaritons appeared\cite{HBP_NP_07}.
In this paper we will discuss how hybrid light-matter excitations in
CCAs can be used to simulate Mott transitions and XX spin models and
how to achieve various quantum information tasks such as quantum
state transfer and cluster state quantum computation
\cite{ASB_PRA_07,AK_NJP_08}.  We note that, intense interest has
arisen since the above early papers that lead to a plethora of
studies on various properties of CCAs in the direction of many body
simulations\cite{greentree-2006,cpsun,fazio,yamamoto,
myungshik-agarwal}, production of photonic and steady state
entanglement\cite{CAB_arxiv_07a,AMB_arxiv_07} and quantum spin
models\cite{HBP_PRL_07,KA_arxiv_08a,CAB_arxiv_08a}.

\section {System Description}
Consider a chain of $N$ coupled cavities\cite{haroche,grangier}. A
realization of this has been studied in structures known as a
coupled resonator optical waveguide (CROWs) or coupled cavity
waveguides (CCW) in photonic crystals, in tapered fibre-coupled
toroidal microcavities  and coupled superconducting microwave
resonators \cite{yariv,orzbay,krauss,trupke}.  The Hamiltonian
corresponds to a series quantum harmonic oscillators coupled through
hopping photons and is given by $H=\sum_{k=1}^{N}\omega_d
a^{\dagger}_{k}a_{k}+\sum_{k=1}^{N}A(a^{\dagger}_{k}a_{k+1}+H.C.)$,
where $a^{\dagger}_{k}(a_{k})$ are the localized eigenmodes (Wannier
functions), i.e.~they describe the creation and annihilation of
photons within individual cavities. The photon frequency and hopping
rate are $\omega_{d}$ and $A$ respectively. There is no
non-linearity present yet since we have not introduced a doping.

The cavities are doped by introducing a single two-level system
(atoms/ quantum dots/superconducting qubits) to each cavity, which,
at site $k$, have ground and excited states $|g\rangle_{k}$ and
$|e\rangle_{k}$ respectively
\cite{coupsupercond,vuckovic,toroid,mabuchi,badolato,supercond1,supercond2,song,kimble-latest}.
The excited state is at an energy $\omega_0$ above that of the
ground state. The resultant Hamiltonian that describes the full
system is the sum of three terms; $H^{free}$ the Hamiltonian for the
free light and dopant parts, $H^{int}$ the Hamiltonian describing
the internal coupling of the photon and dopant in a specific cavity
and $H^{hop}$ for the light hopping between cavities.
\begin{eqnarray}
H^{free}&&=\omega_{d}\sum_{k=1}^N a_k^\dagger a_k+\omega_{0}\sum_k|e\rangle_{k} \langle e|_{k} \\
H^{int}&&=g \sum_{k=1}^N(a_k^\dagger|g\rangle_{k}\langle e|_{k}+H.C.)\\
H^{hop}&&= A\sum_{k=1}^N(a_k^\dagger a_{k+1} +H.C.)
\end{eqnarray}
$g$ is the light atom coupling strength. The $H^{free}+H^{int}$ part of the
Hamiltonian can be diagonalized
 in a basis of mixed photonic and atomic excitations, called
{\it polaritons}. These polaritons, also known as dressed states,
involve a mixture of photonic and atomic excitations and are defined
by the operators $P_{k}^{(\pm,n)}=|g,0\rangle_k \langle n\pm |_{k}$
where $|n\rangle_k$ denote the $n$ photon Fock state in the $k$th
cavity. The polaritons of the $k$th atom-cavity system, denoted by
$|n\pm\rangle_k$, are given by
$|n+\rangle_k=(\sin\theta_{n}|g,n\rangle_k +
\cos\theta_{n}|e,n-1\rangle_k)/\sqrt{2}$ and
$|n-\rangle_k=(\cos\theta_{n}|g,n\rangle_k -
\sin\theta_{n}|e,n-1\rangle_k)/\sqrt{2}$ with energies
$E^{\pm}_{n}=n\omega_{d}\pm g\sqrt{n+\Delta^2/g^2}$,
$\tan(2\theta_{n})=-g\sqrt{n}/\Delta$ and atom-light detuning
$\Delta=\omega_0-\omega_d$. They are also eigenstates of the the sum
of the photonic and atomic excitations operator ${\cal
N}_k=a_k^\dagger a_k+|e\rangle\langle e|_k$ with eigenvalue $n$ (Fig.
1).

\begin{figure}

    \includegraphics[width=0.9\textwidth]{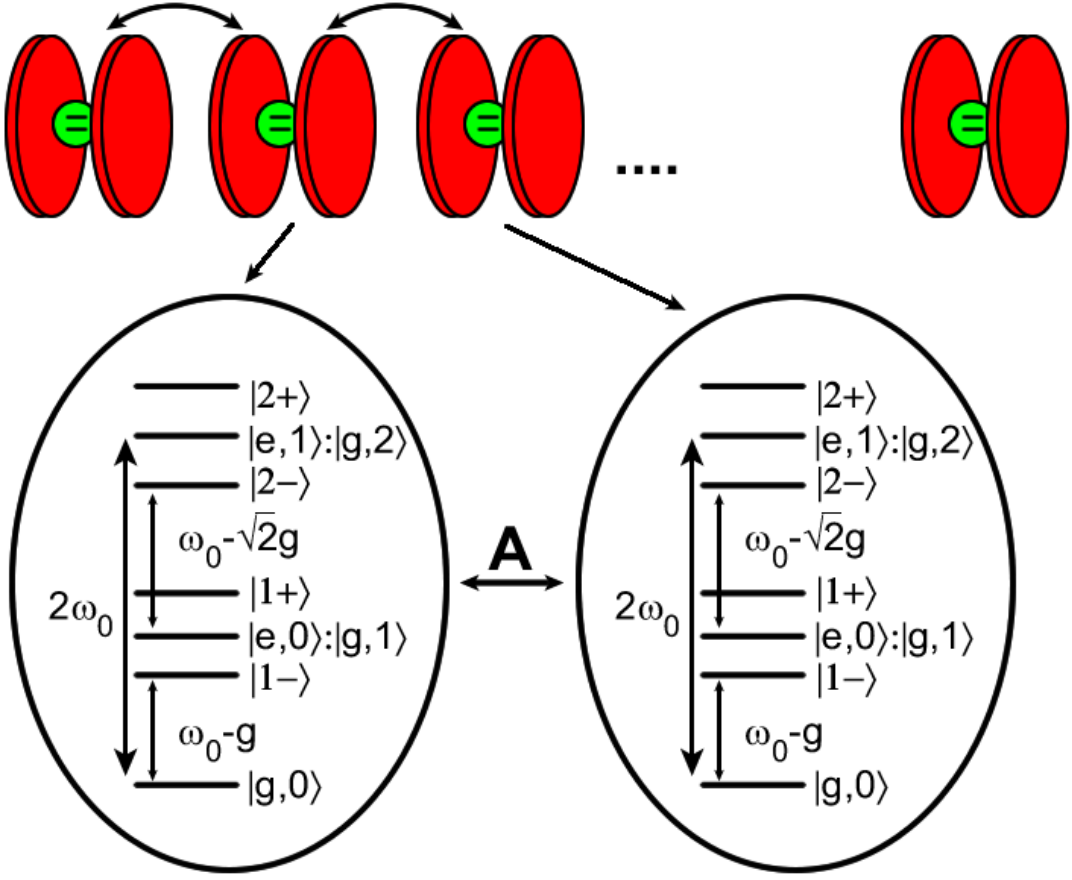}

\caption{A series of coupled cavities coupled through light and the
polaritonic energy levels for two neighbouring cavities. }
\label{pol}
\end{figure}

\subsection{Polaritonic Mott State}

We will now justify that 
the lowest energy states of the system consistent with a given (integer)
number of net excitations per site (or filling factor) becomes a
Mott state of the net (polaritonic) excitations. To understand this, we rewrite the
Hamiltonian (for $\Delta= 0$) in terms of the polaritonic operators
as
\begin{eqnarray}
H=\sum_{k=1}^{N}[\sum_{n=1}^{\infty}n(\omega_{d}-g)P_{k}^{(-,n)\dagger}P_{k}^{(-,n)}+
\sum_{n=1}^{\infty}n(\omega_{d}+g)P_{k}^{(+,n)\dagger}P_{k}^{(+,n)}+\nonumber\\
\sum_{n=1}^{\infty}g(n-\sqrt{n})P_{k}^{(-,n)\dagger}P_{k}^{(-,n)}+
\sum_{n=1}^{\infty}g(\sqrt{n}-n)P_{k}^{(+,n)\dagger}P_{k}^{(+,n)}]+\nonumber\\
A\sum_{k=1}^{N}(a_k^\dagger a_{k+1} +H.C). \label{nodet}
\end{eqnarray}
The above implies (assuming the regime $An<<g\sqrt{n}<<\omega_d$)
that the lowest energy state for a given number, say $\eta$, of net
excitations at the $k$th site would be the state $|\eta-\rangle_k$
(this is because $|\eta+\rangle$ has a higher energy, but same net
excitation $\eta$). Thus one need only consider the first, third and
last lines of the above Hamiltonian $H$ for determining the lowest
energy states. The first line corresponds to a linear spectrum,
equivalent to that of a harmonic oscillator of frequency
$\omega_{d}-g$. If only that part was present in the Hamiltonian,
then it would not cost any extra energy to add an excitation (of
frequency $\omega_{d}-g$) to a site already filled with one or more
excitations, as opposed to an empty site. However, the term
$g(n-\sqrt{n})P_{k}^{(-,n)\dagger}P_{k}^{(-,n)}$ raises energies of
uneven excitation distribution such as
$|(n+1)-\rangle_k|(n-1)-\rangle_{l}$ among any two sites $k$ and $l$
relative to the uniform excitation distribution
$|n-\rangle_k|n-\rangle_{l}$ among these sites. Thus the third line
of the above Hamiltonian can be regarded as an effective non-linear
``on-site" {\it photonic repulsion}, and leads to a Mott state of
the net excitations per site being the ground state for commensurate
filling.  Reducing the strength of the effective non-linearity, i.e.,
the blockade effect through detuning for example, should drive the
system to the superfluid regime. This could be done by Stark shifting the
atomic transitions from the cavity by an external field. The new
detuned polaritons are not as well separated as before and their
energies are merely shifts of the bare atomic and photonic ones by $
\pm g^2n/\Delta$ respectively. In this case it costs no extra energy
to add excitations (excite transitions to higher polaritons) in a
single site, and the system moves to the superfluid regime. Note here that
the mixed nature of the polaritons could in principle allow for
mostly photonic excitations and a photonic Mott state. \
 The required
values of g and $\Delta$ for the corresponding non-linearity though
seem to be unrealistic within current technology
\cite{coupsupercond,vuckovic,toroid,mabuchi,badolato,supercond1,supercond2,song,kimble-latest}.

\begin{figure}

\includegraphics[width=0.8\textwidth]{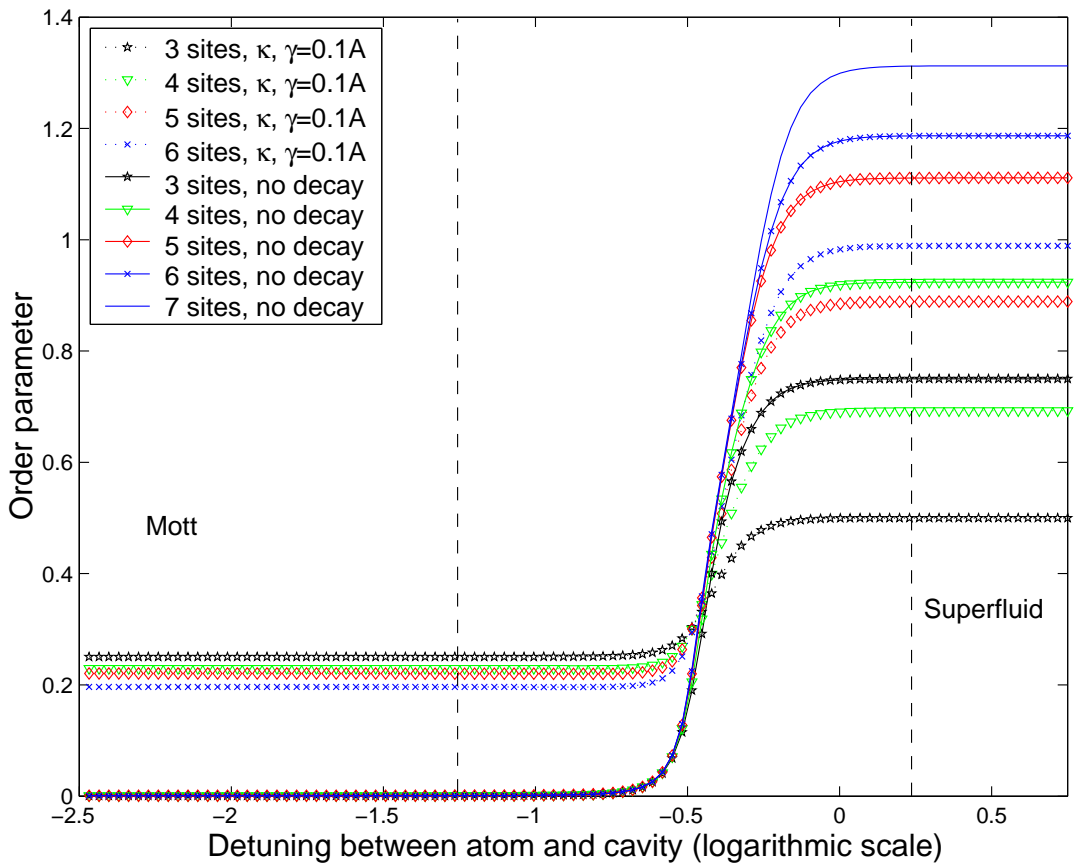}

\caption{The order parameter as a function of the
detuning between the hopping photon and the doped two level
system (in logarithimic units of the matter-light coupling $g$).
Simulations include results for 3-7 sites, with and without
dissipation due to spontaneous emission and cavity leakage. Close to
resonance ($0 \le \Delta/g \le 10^{-1})$, where the photon blockade
induced non-linearity is maximum (and much larger than the hopping
rate), the system is forced into a polaritonic Fock state with the same
integral number of excitations per site (order parameter zero-Mott
insulator state). Detuning the system by applying external fields
and inducing Stark shifts ($\Delta \ge g$), weakens the blockade and
leads to the appearance of different coherent superpositions of
excitations per site( a photonic superfluid). The increase in number of
sites leads to a sharper transition, as expected. } \label{var}
\end{figure}

To quantify the transition of the system from a Mott phase to a
superfluid phase as the detuning $\Delta=\omega_0-\omega_d$ is
increased, we have performed a numerical simulation of the
Hamiltonian of Eqns.(1)-(3) using between 3 and 7 sites (numerical
diagonalization of the complete Hamiltonian without any
approximations) \footnote{We use a finite number of sites for the
simulation as our system compared to the optical lattices case is
computationally more exhaustive. In addition to the bosonic
occupation numbers per site there is also the extra atomic degree of
freedom at each site that cannot be eliminated [R. Roth and K.
Burnett, Phys. Rev. A {\bf 69}, 021601(R) (2004)].}. In the Mott
phase the particle number per site is fixed and its variance is zero
(every site is in a Fock state). In such a phase, the expectation
value of the destruction operator for the relevant particles, the
order parameter, is zero. In the traditional mean field (and thus
necessarily approximate) picture, this expectation value becomes
finite on transition to a superfluid, as a {\em coherent}
superposition of different particle numbers is allowed to exist per
site. However, our entire system is a ``closed" system and there is
no particle exchange with outside. Superfluid states are
characterized by a fixed ``total" number of particles in the finite
site system and the expectation of a destruction operator in any
given site is zero even in the superfluid phase. Thus this
expectation value cannot be used as an order parameter for a quantum
phase transition. Instead we use the variance of total number of
excitations per site, the operator ${\cal N}_k$, in a given site (we
choose the middle cavity, but any of the other cavities would do) to
characterize the Mott to superfluid phase transition. This variance
$var({\cal N}_k)$ has been plotted in Fig.\ref{var} as a function of
$\log_{10}\Delta$ for a filling factor of one net excitation per
site. For this plot, we have taken the parameter ratio $g/A=10^{2}$
($g/A=10^{1}$ gives very similar results), with $\Delta$ varying
from $\sim 10^{-3}g$ to $\sim g$ and $\omega_d,\omega_0 \sim 10^4g$.
We have plotted both ideal graphs (if neither the atoms nor the
cavity fields undergo any decay or decoherence) and also performed
simulations {\em explicitly} using decay of the atomic states and
photonic states in the range of $g/max(\kappa,\gamma)\sim 10^3$,
where $\kappa$ and $\gamma$ are cavity and atomic decay rates.

These decay rates are expected soon to be feasible in toroidal
microcavity systems with atoms \cite{toroid} and arrays of coupled
stripline microwave resonators, each interacting with a
superconducting qubit \cite{supercond1}. For these simulations we
have assumed that the experiment (of going from the Mott state to
the superfluid state and back) takes place in a time-scale of $1/A$ so that
the evolution of one ground state to the other and back is
adiabatic. The simulations of the state with decay have been done
using quantum jumps, and it is seen that there is {\em still a large
difference of $var({\cal N}_k)$ between the Mott and superfluid phases
despite the decays}. As expected the effect of dissipation reduces
the final value of order parameter in the superfluid regime (population has
been lost through decay) whereas in the Mott regime leads to the
introduction of fluctuations, again due to population loss from the
$|1-\rangle$ state. The Mott ($var({\cal N}_k)=0$) to superfluid ($var({\cal
N}_k) > 0$) transition takes place over a finite variation of
$\Delta$ (because of the finiteness of our lattice) around $10g$ and
as expected becomes sharper as the number of sites is increased.

   In an experiment one would start in the resonant (Mott) regime with
all atom-cavity systems initially in their absolute ground states
$(|g,0\rangle^{\otimes k})$ and prepare the atom-cavity systems in
the joint state $|1-\rangle^{\otimes k}$ by applying a global
external laser tuned to this transition. This is the Mott state with
the total (atomic+photonic) excitations operator ${\cal N}_k$ having
the value unity at each site. One would then Stark shift and detune
(globally again) the atomic transitions from the cavity by an
external field and observe the probability of finding each cavity in
$|1-\rangle$ and the predicted decrease of this probability
(equivalent to the increase in our order parameter, the variance of
${\cal N}_k$) as the detuning is increased. For inferring the
fluctuations in ${\cal N}_k$ of our system, it suffices to check the
population of the $|1-\rangle$ state as this is the only way a
single excitation can be present in the $k$th site. For this, a
laser is applied which is of the right frequency to accomplish a
cycling transition between $|1-\rangle$ and another, probe, level
whose fluorescence can be monitored, giving accurate state
measurements \cite{Rowe00}.

\section{Simulating XY spin models}

We will now show that in the
Mott regime the system simulates an XY spin model with the presence
and absence of polaritons corresponding to spin up and down. Let us
assume we initially populate the lattice only with polaritons of
energy $\omega_{0}-g$. In the limit $\omega_{d}\approx\omega_{0}$,
Eqs. (4) becomes
\begin{eqnarray}
H_{k}^{free}=\omega_{d}\sum_{k=1}^{N}&&P^{(+)\dagger}_{k}P^{(+)}_{k}+
P^{(-)\dagger}_{k}P^{(-)}_{k}\\
H_{k}^{int}=g\sum_{k=1}^{N}&&P^{(+)\dagger}_{k}P^{(+)}_{k}-P^{(-)\dagger}_{k}P^{(-)}_{k}\\
H_{k}^{hop.}=A\sum_{k=1}^{N}&&P^{(+)\dagger}_{k}P^{(+)}_{k+1}+P^{(-)\dagger}_{k}P^{(+)}_{k+1}
+\nonumber\\
&&P^{(+)\dagger}_{k}P^{(-)}_{k+1}+P^{(-)\dagger}_{k}P^{(-)}_{k+1}+H.C.
\label{H_hop_pol}
\end{eqnarray}
where $P_{k}^{(\pm)\dagger}=P_{k}^{(\pm,1)\dagger}$ is the
polaritonic operator creating excitations to the first polaritonic
manifold (Fig. 1). In the rotating wave approximation, Eq.
\ref{H_hop_pol} reads (in the interaction picture).
$H_{I}=A\sum_{k=1}^{N}P^{(-)\dagger}_{k}P^{(-)}_{k+1}+H.C.$
In deriving the above, the logic requires two steps. Firstly note
that the terms of the type $P^{(-)\dagger}_{k}P^{(+)}_{k+1}$, which
inter-convert between polaritons, are fast rotating and they vanish.
Secondly, if we create only the polaritons $P^{(-)\dagger}_{k}$ in
the lattice, then the polaritons corresponding to
$P^{(+)\dagger}_{k}$ will never even be created, as the
inter-converting terms are vanishing. Thus the term
$P^{(+)\dagger}_{k}P^{(+)}_{k}$ can also be omitted. Note that
because the double occupancy of the sites is prohibited, one can
identify $P^{(-)\dagger}_{k}$ with
$\sigma^{+}_k=\sigma^x_k+i\sigma^y_k$, where $\sigma^x_k$ and
$\sigma^y_k$ are standard Pauli operators. Then the Hamiltonian
becomes $ H_I=\sigma^x_k\sigma^x_{k+1}+\sigma^y_k\sigma^y_{k+1}$
which is the standard XY model of interacting spins with spin
up/down corresponding to the presence/absence of a polariton. Note
that although this is different to optical lattice realizations of
spin models, where instead, the internal levels of a two level atom
are used for the two qubit states \cite{Duan-Lukin-Demler-PRL}, the
measurement could be done using very similar atomic state
measurement techniques (utilizing the advantage of larger distances
between sites here).

\section{Cluster state quantum computation}
{\em Cluster state generation:} The typical implementation of
cluster state quantum computing\cite{cluster1,clusterall}, requires
initializing all qubits in a 2D lattice in the
$\ket{+}=(\ket{0}+\ket{1})/\sqrt{2}$ state and then performing
controlled-phase gates ($CP$) between nearest-neighbours. In the
present CCA system, we have no direct two-qubit gate and the
available interaction is not of the Ising type, which
straightforwardly gives controlled-phase gates, but an `always on'
global Hamiltonian coupling of the XY form. Some consideration of
similar scenarios has previously been made \cite{loss}, although
these have primarily concentrated on the Heisenberg interaction. In
comparison, the technique which we invoke induces entanglement in a
more stable way (from the exchange of two effective fermions
\cite{Christandl:2004a,Jaksch:2004a}, and hence it is topological in
nature), requires fewer control structures but is inapplicable to
the case of Heisenberg coupling. Moreover, the strategy that we will
outline momentarily is specifically designed to cope with the
always-on nature of the interaction -- this is an aspect which is
often neglected when forming a cluster state either from Hamiltonian
interactions such as the Ising model or as the ground state of a
Hamiltonian \cite{rudolph:06}; one must disable the system dynamics
once the state has been formed.

\begin{figure}
\includegraphics[width=0.7\textwidth]{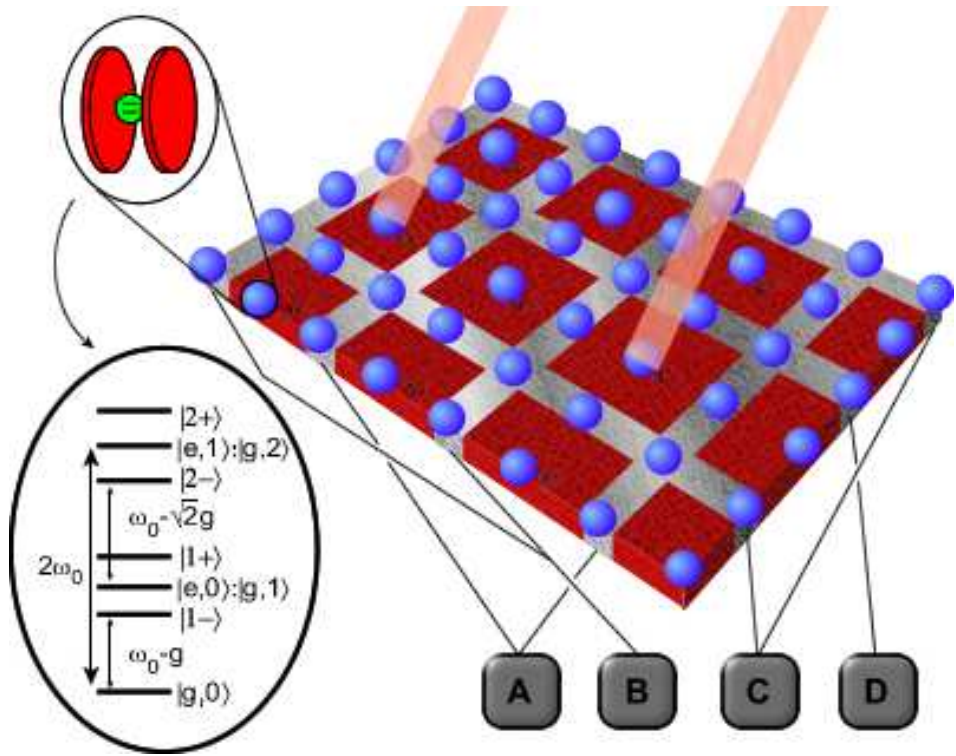}
\caption{We work with a 2D array of atom-cavity systems. When the
atom is on resonance with the cavity, the ground state $|g,0\rangle$
and the first excited state $|1-\rangle$ of the combined atom-photon
(polaritonic) system in each site can be used as qubits as no other
states are accessible\cite{ASB_PRA_07}. By applying Stark shifts
with control electrodes or properly tuned laser fields to sets of
qubits (the gates shown under the qubits), we disable the exchange
Hamiltonian of a qubit to all of its neighbours creating isolated
chains of three qubits. Within each chain, the two extremal qubits
are the computational qubits, and the central qubit acts as a
mediator. Using only four different groupings of three-qubit chains,
we can generate a cluster state. Individual single qubit rotations
and measurements are possible and made by properly applying local
external fields utilizing the fact that the cavities can be well
separated. } \label{pol}
\end{figure}
 This requirement can be realized by combining the
system's natural dynamics with a protocol where some of the
available physical qubits are allocated as gate ``mediators"  and
the rest as the logical qubits. The mediator atoms can be Stark
shifted on and off resonance from their cavities through the
application of an external field, inhibiting the photon hopping and
thereby isolating each logical qubit. The same inhibition of
couplings will be used to generate the cluster state. We note here
that the error introduced in the step is due to a second-order
transition between on-resonance qubits (via a dark-passage through
the central off-resonant qubit), which is thus suppressed by a
factor of order $A/\Delta$, where $\Delta=\omega_d-\omega_0$ is the
detuning of the off-resonant cavity.

Before describing the 4-step global gate sequence to create the
cluster state, first observe that to generate the control phase, it
is enough to localize chains of 3 qubits, let them evolve for a time
$t_0=\pi/(2\sqrt{2}A)$ and then apply a measurement on the middle
`mediator' qubit (in the $\sigma^z$ basis). Depending on the
measurement result, $\ket{0}$ or $\ket{1}$, a non-local gate is
generated between the remaining two qubits, either {\sc
SWAP}.$(\sigma^z\otimes\sigma^z).CP$ or {\sc SWAP}.$CP$ respectively
\cite{Christandl:2004a}. In both cases, the gates in addition to the
$CP$ are Clifford gates, and can thus be recorded and taken into
account during the measurement-based computation with the help of an
efficient classical computation. Alternatively, if the mediator
starts in a known state, say $\ket{0}$, then measuring it and
post-selecting on the $\ket{0}$ outcome acts as a useful form of
error suppression against timing errors (perfectly) and some forms
of decoherence (giving some improvement). On failure (the $\ket{1}$
result), we can make use of the techniques from Benjamin et al.
\cite{clusterall} to fix the error.

Our sequence to generate the cluster state is initiated by preparing
all qubits in the $\ket{+}$ state through the application of global
$\pi/2$ pulse. One quarter of the sites will be used as logical
qubits and the rest as ``mediators" and ``off" qubits
interchangeably. By tuning qubits ``off" (i.e.~moving them off
resonance with the aid of the Stark effect, creating an energy cost
for photon hopping), we control their interaction with their nearest
neighbours and separate the array into chains of three qubits. The
steps to create the cluster states are as follows (Fig.~\ref{pol}).
First apply gates B, C, D, which take the corresponding systems
addressed by them off resonance. By doing this, groups of three `on'
qubits are created and isolated from each other. For these we apply
the `three qubit' protocol described above and  a $CP$ is performed
between the extremal qubits of each group. Now for every second line
we have every second qubit `C-Phased'. In the next step we will
connect these pairs to each other by applying A, C, D and the $CP$
protocol again (interchanging the role of previously `off' qubits
with mediators). After this stage we have successfully prepared
complete rows of qubits in the cluster state. Now we need to connect
the columns, which is done by applying the A, B and C gates along
with the $CP$ part. Finally, by applying A, B and D, those pairs of
columns can be connected, leading to a 2D cluster for every second
qubit in the whole array. The required measurement sequence for a
particular algorithm is then applied, utilizing the local
accessibility of the sites (in any implementation these qubits are
at least a few micrometers apart)
\cite{coupsupercond,vuckovic,toroid,mabuchi,
badolato,supercond1,supercond2,song,kimble-latest}.

As outlined, the initial state of the mediator qubit is irrelevant
to the success of the scheme (provided it is in the $\ket{g,0},
\ket{1-}$ subspace) due to our measurement of it. This protects
against decoherence while the atom-cavity system is off resonance.
If, instead, we knew that the initial state of the mediator qubit
was $\ket{0}$, say, then the measurement would provide a mechanism
for low-level error suppression during the application of the gate
(projection onto the $\ket{0}$ state leads to the Zeno effect) and
detection (if the measurement result is $\ket{1}$). Providing the
error probability is small, where small is related to the
percolation threshold of the system \cite{cluster1}, knowing where
these errors occurred allows one to route the computation around the
defects. For noisier systems, other techniques can be explored
\cite{kieling-2006}.

\subsection{Consideration of Errors}

\begin{figure}
\includegraphics[width=0.6\textwidth]{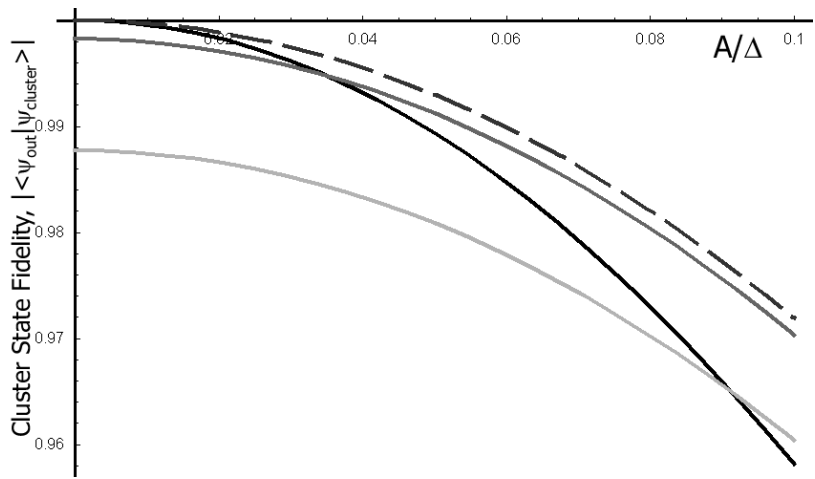}
\caption{The fidelity of generation of a cluster state on a 3x3 grid
of cavities, as the detuning $\Delta$ of the mediator off-resonance
cavities is varied (in units of the hopping A). The dashed line
includes post-selection on getting $\ket{0}$ outcomes when measuring
off-resonance qubits. The grey lines also incorporate spontaneous
decay and cavity leakage of $0.05A$ (dark) and $0.08A$ (light).}
\label{simulation} \vspace{-0.5cm}
\end{figure}

Aside from the aforementioned effect the comes through second-order
perturbation theory, which is the primary assumption we have made in
deriving the system dynamics, and which results in an error of order
$A/\Delta$, what other practical concerns are likely to limit the
usefulness of our scheme? Primarily, our concern should be
decoherence, which will typically manifest as cavity leakage and
spontaneous emission from the atoms. In Fig.~\ref{simulation}, we
calculate the fidelity of generation of a cluster state on a 3x3
grid of cavities, as the detuning $\Delta$ of the mediator
off-resonance cavities is varied. The dashed line includes
post-selection on getting $\ket{0}$ outcomes when measuring
off-resonance qubits, while grey lines also incorporate spontaneous
decay and cavity leakage. We observe that the fidelity remains
larger than $0.97$ even when relatively large values of dissipation
are included. More sophisticated schemes have the potential to
further reduce the experimental errors. For example, standard
Hamiltonian simulation techniques allow us to negate the second
order exchange term due to the off-resonance cavities, simply by
repeatedly applying $\sigma_z$ gates to every second on-resonance
triplet throughout the evolution. One might even hope that we could
use this coherent effect to enhance the scheme through the use of,
for example, optimal control techniques.
Most of the errors considered here (cavity leakage, spontaneous
emission of the atom, and on-off detuning of qubits) are local
effects, introducing local noise, which can ultimately be addressed
by fault-tolerant techniques \cite{raussendorf}.

Another class of properties that could be expected to have an effect
are timing errors (when the external fields are applied, and how
quickly they can be ramped up to maximum strength), and problems
with system identification or manufacture. If the system is
improperly identified or manufactured, then we will be using an
incorrect timescale for the evolution and, as such, it is equivalent
to a timing error. Within the difficulties of imperfect system
manufacture is the problem of ensuring that the atoms and cavities
are on-resonance. However, if they are slightly off resonance, and
we can determine this, external fields can be used to compensate. If
this is not possible, then, in fact, it does not cause a problem
provided the detuning is sufficiently small that we are still within
the Mott insulator phase \cite{ASB_PRA_07}, the only difference will
be a slight change in the effective coupling between cavities, and
hence another timing effect. Thanks to the mediator spin,
specifically our ability to measure it, we have a geometric
robustness to timing errors \cite{Kay:2005b} i.e.~if our timing
error is $O(\delta t)$, the accuracy with which the evolution is
achieved is only faulty by $O(\delta t^2)$. Finally, the entangling
operation, which is the essential part of the whole scheme, has a
topological robustness -- tuning the parameters of the Hamiltonian
differently leaves the generated phase entirely unaffected provided
the evolution has completed successfully. Essentially this is a
result of the fact that the presence of $\ket{1}$s in the system can
be mapped to the presence of fermions, and it is the topological
robustness of the $-$ve sign appearing when two fermions exchange
which we are using \cite{Christandl:2004a}.
\subsection {Implementing algorithms: } Initial experimental
algorithmic implementations with coupled cavities can be expected to
utilize the most basic building block of our scheme, a $3\times 3$
grid of cavities, which allows us to generate a four-qubit cluster
state. As with the four-photon cluster state initially used by
Walther {\em et al.} and more recently by Pan {\em et al.},
\cite{Walther}, this cluster state would be suitable for
demonstrating the preparation of an arbitrary one-qubit state, an
entangling gate between two qubits, and even the implementation of
Grover's search algorithm on two qubits \cite{Pan}. For example, by
applying the local gates $H\otimes H\otimes\sigma_z\otimes\sigma_z$,
where $H$ is the Hadamard rotation, we convert of `box' cluster that
the $3\times 3$ grid prepares into the 1D cluster state of 4 qubits,
which is given the interpretation of a single qubit, and
measurements on the state yield quantum gates on this single qubit.
Moreover, generation of this four qubit cluster state is simpler
than generation of an arbitrarily sized cluster state because we
only need two control steps instead of four, thereby keeping us even
further within the decoherence time of the system.

\begin{figure}
\includegraphics[width=0.7\textwidth]{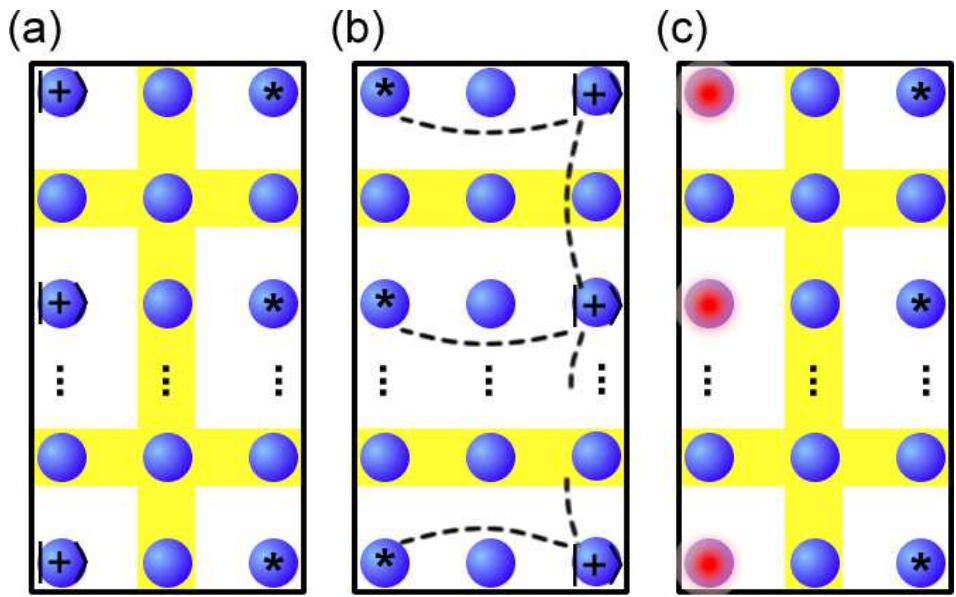}
\caption{ Sequence for minimising the number of qubits required for
a cluster state computation. (a) After the first $n-1$ steps of the
algorithm, the first column of qubits is initialised in the
$\ket{+}$ state, and the third column, with qubits denoted by $*$,
are in the state of output for the first $n-1$ steps of the
computation. (b) We use control sequences, bringing mediator qubits
on resonance, to convert the $\ket{+}$ states into a cluster state,
and to entangle them with the output qubits. The SWAP in the
entangling operation moves these output qubits to the first column.
(c) Measure the qubits of the first column as corresponds to the
$n^{th}$ step of the computation, and reinitialise in the $\ket{+}$
state. The rightmost column corresponds to the output. The sequence
then repeats.} \label{panels} \vspace{-0.5cm}
\end{figure}

Perhaps the next important step would then be to demonstrate Shor's
factoring algorithm, the factoring of 15 being the standard
demonstration. To implement as a cluster state computation, the six
computational qubits \cite{shor_imp} translate into the requirement
of a cluster state that is eleven qubits wide. Hence, we need an
array which is 21 cavities wide. The breadth of the cluster state,
which corresponds to time in the circuit model, is a quantity that
we can trade against the time taken for the computation. At one
extreme, we can create the whole cluster state in one go, with the
simple set of four steps already outlined, and we benefit from the
large degree of parallelism available to us. This requires a 2D grid
of cavities of size $21\times 311$ \footnote{To arrive at this
required number of gates, we have taken the circuit presented in
\cite{shor_imp} and converted it into a nearest-neighbor, 2-qubit
gate algorithm. Hence, the possibility for some small degree of
optimisation in the number of qubits remains.}. At the other
extreme, a grid of $21\times 3$ cavities suffices. In this case, one
starts with the $11\times 2$ cluster state, and performs one time
step of measurement (i.e.~measure the 11 qubits in one column). The
result remains in the other column. We then repeat the cluster state
generation process, reinitialising the measured qubits in the
cluster state, and performing the next time step
(Fig.~\ref{panels}). This requires 156 consecutive entangling steps,
but the reinitialising of the cluster state after measurement
eliminates the effect of decoherence over this timescale. Any
combination between these two extremes is also possible, and is a
necessary property of any scalable implementation of cluster state
computation for the sake of preventing decoherence.

Once initial cluster state experiments have been performed, it
simply becomes a question of how many cavities one can reasonably
couple together. Alternatively, since the two-qubit gate that we can
generate is entangling (and hence universal for quantum
computation), we can also consider using it directly to implement
the circuit model of computation. This has a much smaller overhead
of qubits, but instead requires much higher quality cavities. For
example, to factor 15 we would only need a $5\times 3$ grid of
cavities to give us six computational qubits. However, we would need
approximately 15 consecutive entangling steps (we have attempted to
minimise this number by allowing as many of the gates to be applied
in parallel as possible, and by optimising the initial labelling of
each qubit), hence requiring a time of order $15\pi/(\sqrt{2}A)$.
Hence, to reduce the effect of dissipative decay, we require an
order of magnitude improvement in the decoherence properties of the
qubits to compensate for the increased running time.

\section{ Experimental implementations }

As previously mentioned, there are three primary candidate
technologies; fibre coupled micro-toroidal cavities, arrays of
defects in PBGs  and superconducting qubits coupled through
microwave stripline resonators
\cite{coupsupercond,vuckovic,toroid,mabuchi,
badolato,supercond1,supercond2,song,kimble-latest}. In order to
achieve the required limit of no more than one excitation per site
\cite{ASB_PRA_07}, the ratio between the internal atom-photon
coupling and the hopping of photons down the chain should be of the
order of $g/A\sim10^{2}-10^{1}$($A$ can be tuned while fabricating
the array by adjusting the distance between the cavities and $g$
depends on the type of the dopant). In addition, the cavity/atomic
frequencies should be $\omega_d,\omega_0 \sim 10^{4}g,10^{5}g$ and
the losses should also be small, $g/\max(\kappa,\gamma)\sim 10^3$,
where $\kappa$ and $\gamma$ are cavity and atom/other decay rates.
The polaritonic states under consideration are essentially
unaffected by decay for a time $10/A$ ($10$ns for the toroidal case
and $100$ns for microwave stripline resonators). While the decay
time of $10/A$ may seem uncomfortably close to the preparation time
for a cluster state, $\sqrt{2}\pi/A$, the previously described
technique (Fig.~\ref{panels}) of continuously reforming the cluster
state and connecting it to the output of the previous stage allows a
continuous computation that exceeds the decay time for an individual
cavity. The required parameter values are currently on the verge of
being realised in both toroidal microcavity systems with atoms and
stripline microwave resonators coupled to superconducting qubits,
but further progress is needed. Arrays of defects in PBGs remain one
or two orders of magnitude away, but recent developments, and the
integrability of these devices with optoelectronics, make this
technology very promising as well. In all implementations the cavity
systems are well separated by many times the corresponding
wavelength of any local field that needs to be applied in the system
for the measurement process.

\section{Conclusions}

In this paper, we showed that a range of many-body system effects,
such Mott transitions for polaritonic particles obeying mixed
statistics could be observed in optical systems of arrays
individually addressable coupled cavities interacting with two level
systems. We also showed the capability and advantages of simulating
XY spin models using our scheme and noted the ability of these
arrays to simulate arbitrary quantum networks . In addition we
discussed how universal quantum computation could be realized in a
coupled array of individually addressable atom-cavity systems, where
the qubits are given by mixed light-matter excitations in each
cavity site. While single-qubit operations can be locally achieved,
the only available interaction between qubits is due to the natural
system Hamiltonian. We show how to manipulate this to give a
controlled-phase gate between pairs of qubits. This allows
computation either using the circuit model, or a measurement-based
computation, the latter being most suited to reducing experimental
errors. We have discussed possible architectures for implementing
these ideas using photonic crystals, toroidal microcavities and
superconducting qubits and point out their feasibility and
scalability with current or near-future technology. We also
discussed possible implementations using photonic crystals, toroidal
microcavities and superconducting systems.

 We acknowledge the hospitality of the Centre for Quantum
Technologies in NUS, Singapore. This work was supported in part by
the E.U. FP6-FET Integrated Project SCALA (CT-015714) and the
National Research Foundation $\&$ Ministry of Education, Singapore.

\end{document}